\documentclass[12pt, a4paper]{article}
\usepackage{amsmath,amssymb}
\usepackage[dvips]{graphicx}
\usepackage{color}

\begin{document}

\begin{titlepage}

\begin{center}

\hfill UT-17-15\\

\vskip .75in

{\Large \bf 
Can decaying particle explain cosmic infrared \\ \vspace{2mm}background excess?
}

\vskip .75in

{\large Kazunori Kohri$^{a,b}$, Takeo Moroi$^{c,d}$ and Kazunori Nakayama$^{c,d}$}

\vskip 0.25in

\begin{tabular}{ll}
$^{a}$ &\!\! {\em Institute of Particle and Nuclear Studies, KEK, }\\
& {\em 1-1 Oho, Tsukuba, Ibaraki 305-0801, Japan}\\[.3em]
$^{b}$ &\!\! {\em The Graduate University for Advanced Studies (SOKENDAI),}\\
&{\em 1-1 Oho, Tsukuba, Ibaraki 305-0801, Japan}\\[.3em]
$^{c}$ &\!\! {\em Department of Physics, Faculty of Science, }\\
& {\em The University of Tokyo,  Bunkyo-ku, Tokyo 113-0033, Japan}\\[.3em]
$^{d}$ &\!\! {\em Kavli IPMU (WPI), UTIAS,}\\
&{\em The University of Tokyo,  Kashiwa, Chiba 277-8583, Japan}\\[.3em]

\end{tabular}

\end{center}
\vskip .5in

\begin{abstract}
  Recently the CIBER experiment measured the diffuse cosmic infrared
  background (CIB) flux and claimed an excess compared with integrated
  emission from galaxies.  We show that the CIB spectrum can be
  fitted by the additional photons produced by the decay of a new
  particle.  However, it also contributes too much to the anisotropy of the CIB,
  which is in contradiction with the anisotropy measurements by the CIBER and Hubble Space Telescope.
\end{abstract}

\end{titlepage}


\renewcommand{\thepage}{\arabic{page}}
\setcounter{page}{1}
\renewcommand{\thefootnote}{\#\arabic{footnote}}
\setcounter{footnote}{0}

\newpage

\section{Introduction}

Origins of the total flux of the diffuse cosmic infrared background
(CIB) radiation have not been known for certain. However, it is 
 believed that the CIB should be integrated radiation of photons
emitted in the past Universe through the cosmic history. For
example, the infrared radiation can be produced at least in standard
mechanisms of galaxy formations and
evolutions~\cite{Franceschini:2008tp}. Therefore, precise observational
data are expected to constrain unknown mechanisms in
which the CIB are additionally produced, e.g., as relics of
redshifted ultraviolet photons emitted at an epoch of cosmic
reionization induced by Pop III stars, young galaxies, or black
holes~\cite{Santos:2002hd,Salvaterra:2002rg,Fernandez:2005gx,Mii:2005as,Yue:2012dd,Yue:2013hya}.

Quite recently the Cosmic Infrared Background Experiment (CIBER)
collaboration reported their first result of spectral measurements for
the diffuse CIB radiation in unexplored wavelength ranges from 
0.8\,$\mu$m to 1.7\,$\mu$m by their sounding rocket
experiments~\cite{Matsuura:2017}. Remarkably, they found that there
exist significant excesses in their data compared with the previous
prediction deduced theoretically by indirect observations with
counting numbers of galaxies~\cite{Franceschini:2008tp}.  This
discrepancy cannot be resolved even if they adopt a model-independent
lower limit on the observed CIB radiation~\cite{Matsuura:2017}. So far
there are no natural astrophysical candidates which explains those
excesses of the CIB radiation at around the wavelength $\sim$ 1
$\mu$m.

In this paper, we study scenarios in which the CIB spectrum is fitted
by additional photons produced by a decaying hypothetical particle. 
However, such a scenario predicts too large CIB anisotropy spectrum,
which contradicts with the measurement by the Hubble Space telescope (HST)~\cite{Mitchell-Wynne:2015rha}
and CIBER itself~\cite{Zemcov:2014eca}.

The plan of the paper is as follows. 
In Sec.~\ref{sec:2} we discuss the possibility that the decaying particle explains the excess CIB radiation
and constraints from star cooling.
In Sec.~\ref{sec:3}, the CIB anisotropy spectrum in the decaying particle model is studied.
It is shown that models to explain the excess in the mean CIB intensity contradicts with the 
measurement of the CIB anisotropy spectrum.
Sec.~\ref{sec:4} is devoted to conclusions and discussion.

\section{CIB from Decaying Particle}
\label{sec:2}

\subsection{Mean Intensity from Decaying Particle}

Let us suppose that there is a light particle $\phi$ which decays into
another particle $\chi$ plus photon $\gamma$.  For the moment we do
not specify the model: $\phi$ and $\chi$ can either be bosons or
fermions, which are assumed to be singlets under the standard-model
gauge group.

Independently of the detailed model, the CIB spectrum from the $\phi$ decay can be calculated by the following three parameters:
the photon energy produced by the decay ($E_{\rm max}$), 
the lifetime of $\phi$ ($\tau_\phi$) and the abundance of $\phi$ ($Y_\phi$).
Here $Y_\phi = n_\phi/s$ denotes the number-to-entropy density ratio of $\phi$.\footnote{
	If $\tau_\phi \lesssim t_0$ ($t_0$ denotes the present age of the universe), $n_\phi/s$ is not regarded as a constant.
	We define $Y_\phi$ as the number-to-entropy density ratio at the high-enough redshift: $Y_\phi \equiv (n_\phi/s)_{t\ll \tau_\phi}$.
}
From the kinematics, the photon energy produced by the decay is given by
\begin{align}
	E_{\rm max} =\frac{m_\phi^2-m_\chi^2}{2m_\phi},
\end{align}
where $m_\phi$ and $m_\chi$ are masses of $\phi$ and $\chi$, respectively.
For later use we also define $r_\phi$, the energy fraction of $\phi$ in the total dark matter density $\rho_{\rm DM}$, 
\begin{align}
	r_\phi \equiv \frac{m_\phi Y_\phi}{\rho_{\rm DM}/s}.
\end{align}
Thus $Y_\phi$ is determined by the combination $m_\phi/r_\phi$.
Later we will calculate the CIB spectrum by varying $E_{\rm max}$, $\tau_\phi$ and $m_\phi/r_\phi$.

Since the produced photon energy is monochromatic, the present photon energy $E$ has one-to-one correspondence to the 
redshift at the injection $z$ as $E(z) = E_{\rm max}/(1+z)$.
Neglecting the intergalactic/interstellar absorption of photons and considering photons produced after the matter-domination, 
the present CIB energy flux in the flat universe is given by
\begin{align}
	I(E)= E^2\int dr W(r) =\frac{c}{4\pi} \frac{n_\phi(z) a^3(z)}{\tau_\phi} \frac{E}{H(z)},  \label{flux}
\end{align}
where $c$ is the speed of light, $a(z)=(1+z)^{-1}$ is the cosmic scale factor,
$H(z)$ is the Hubble parameter at the redshift $z$, which is related to the present Hubble parameter $H_0$ through
\begin{align}
	H(z) = H_0\left(\Omega_\Lambda + \Omega_m(1+z)^3 \right)^{1/2},
\end{align}
with $\Omega_\Lambda$ and $\Omega_m$ being the dark energy and dark matter density parameter and
\begin{align}
	W(r) \equiv \frac{c}{4\pi} \frac{n_\phi(z) a^3(z)}{\tau_\phi} \frac{dN_\gamma}{dE'},   \label{Wr}
\end{align}
with $r(z)$ being the comoving distance to the redshift $z$,
\begin{align}
	r(z) = \int_0^z \frac{c\,dz}{H(z)},
\end{align}
and the photon spectrum at the injection has a delta-function shape in our case:
\begin{align}
	\frac{dN_\gamma}{dE'} = \delta(E'-E_{\rm max}),
\end{align}
with $E'=(1+z)E$.
Assuming $\phi$ particles are non-relativistic, we can analytically express its number density at the redshift $z$ as
\begin{align}
	n_\phi(z) = Y_\phi\,s(z) \exp\left(-\frac{t(z)}{\tau_\phi}\right),
\end{align}
where the cosmic time at the redshift $z$ is given by
\begin{align}
	t(z) = \frac{1}{3H_0 \sqrt{\Omega_\Lambda}}\ln\left( \frac{\sqrt{\Omega_\Lambda + \Omega_m(1+z)^3}+\sqrt{\Omega_\Lambda}}{\sqrt{\Omega_\Lambda + \Omega_m(1+z)^3}-\sqrt{\Omega_\Lambda}}\right),
\end{align}
if the radiation energy density is negligible.
The peak energy in the photon spectrum is then found to be
\begin{align}
	E_{\rm peak} \simeq \begin{cases}
		E_{\rm max}  & {\rm for}~~\tau_\phi \gtrsim t_0 \\
		E_{\rm max}\left( \frac{3H_0 \tau_\phi\sqrt{\Omega_m}}{2} \right)^{2/3} & {\rm for}~~\tau_\phi \lesssim t_0
	\end{cases}.
\end{align}
An inspection shows that the intensity scales as $I(E) \propto E^{5/2}$ for $E \ll E_{\rm peak}$.

\begin{figure}[t]
\begin{center}
\includegraphics[scale=1.2]{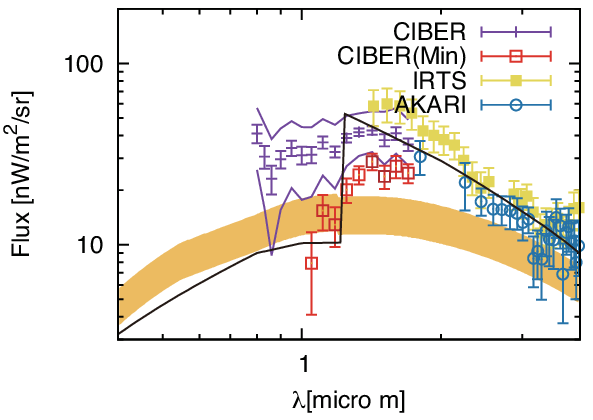}
\includegraphics[scale=1.2]{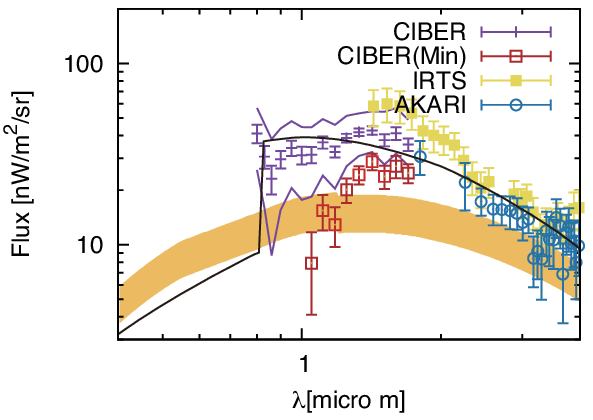}
\includegraphics[scale=1.2]{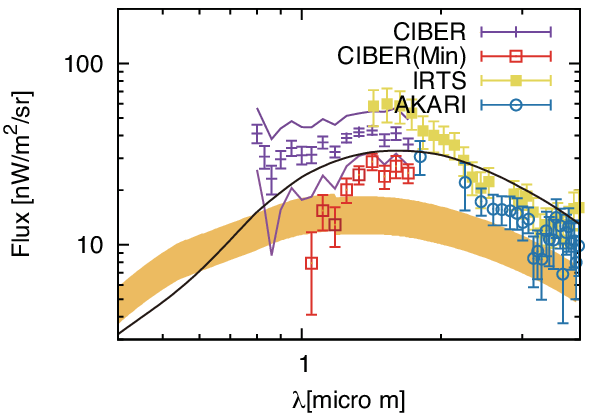}
\caption{
The CIB photon spectrum from decaying particle.
Data points include CIBER with Kelsall ZL model, CIBER with minimum EBL model, AKARI and IRTS
with only statistical error bars. For the CIBER data with Kelsall ZL model, we also show the range of systematic error.
The orange band shows the inferred CIB level by the HESS telescope.
The solid line corresponds to the contribution from decaying particle (plus expected background from galaxy counts) for the following parameter set: 
$E_{\rm max}=1$\,eV, $m_\phi/r_\phi=3$\,eV, $\tau_\phi=2\times10^{22}$\,sec (upper left),
$E_{\rm max}=1.5$\,eV, $m_\phi/r_\phi=90$\,keV, $\tau_\phi=2.5\times10^{17}$\,sec (upper right)
and $E_{\rm max}=8$\,eV, $m_\phi/r_\phi=80$\,keV, $\tau_\phi=1\times10^{16}$\,sec (lower). 
}
\label{fig:spec}
\end{center}
\end{figure}

Fig.~\ref{fig:spec} shows the resulting photon spectrum for several parameter choices.
As for cosmological parameters, we take the best-fit values obtained by the Planck satellite~\cite{Ade:2015xua}.
Data points include CIBER~\cite{Matsuura:2017} with Kelsall zodiac light (ZL) model~\cite{Kelsall:1998bq}, 
CIBER with minimum extragalactic background light (EBL) model~\cite{Matsuura:2017}, AKARI~\cite{Tsumura:2013iza}
and IRTS~\cite{Matsumoto:2015fma} with only statistical error bars. 
For the CIBER data with Kelsall ZL model, we also show the range of systematic error.
The orange band shows the inferred CIB level from TeV gamma-ray measurements from distant sources by the HESS telescope~\cite{Abramowski:2012ry}.
The solid line shows the contribution from decaying particle (plus expected background from galaxy counts~\cite{Gilmore:2011ks}) for the following parameter set: 
$E_{\rm max}=1$\,eV, $m_\phi/r_\phi=3$\,eV, $\tau_\phi=2\times10^{22}$\,sec (upper left),
$E_{\rm max}=1.5$\,eV, $m_\phi/r_\phi=90$\,keV, $\tau_\phi=2.5\times10^{17}$\,sec (upper right)
and $E_{\rm max}=8$\,eV, $m_\phi/r_\phi=80$\,keV, $\tau_\phi=1\times10^{16}$\,sec (lower).
For the first parameter set, the lifetime is much longer then the age of the universe and the spectrum shows a sharp cutoff.
It is roughly consistent with AKARI and IRTS data at longer wavelength and CIBER with minimum EBL at shorter wavelength.
For the second/third parameter set, a significant fraction of $\phi$ already decayed at the present epoch.\footnote{
	Since masses of $\phi$ and $\chi$ are strongly degenerate, the energy density of $\phi$ (or $\chi$) does not change
	due to the decay.
}
It is consistent with AKARI and IRTS data at longer wavelength and CIBER with Kelsall ZL model at shorter wavelength.

As seen from this figure, taking account of uncertainties of the measured CIB spectrum, 
relatively broad parameter region is allowed to explain the CIB excess.
Several comments and constraints on the parameters are listed below.
\begin{itemize}
\item
If $\tau_\phi\gtrsim t_0$, the overall flux is determined by the combination $r_\phi / (m_\phi \tau_\phi)$,
while the $\phi$ energy density should be smaller than that of  dark matter: $r_\phi \leq 1$
and also we must have $m_\phi \geq 2E_{\rm max}$.
Under these conditions the lifetime cannot be longer than $2\times 10^{22}$\,sec to explain the CIB excess.

\item
If $E_{\rm max}$ is larger than 13.6\,eV, photons produced before the reionization epoch contribute to the extra ionizing source.
Possible injected energy is severely constrained in such a case~\cite{Chen:2003gz}.
Since we need $E_{\rm peak}\sim 1$\,eV to fit the CIBER/IRTS data, it gives lower bound on the lifetime as $\tau_\phi \gtrsim 10^{16}$\,sec.
\end{itemize}

\begin{figure}[t]
  \begin{center}
    \includegraphics[scale=0.8]{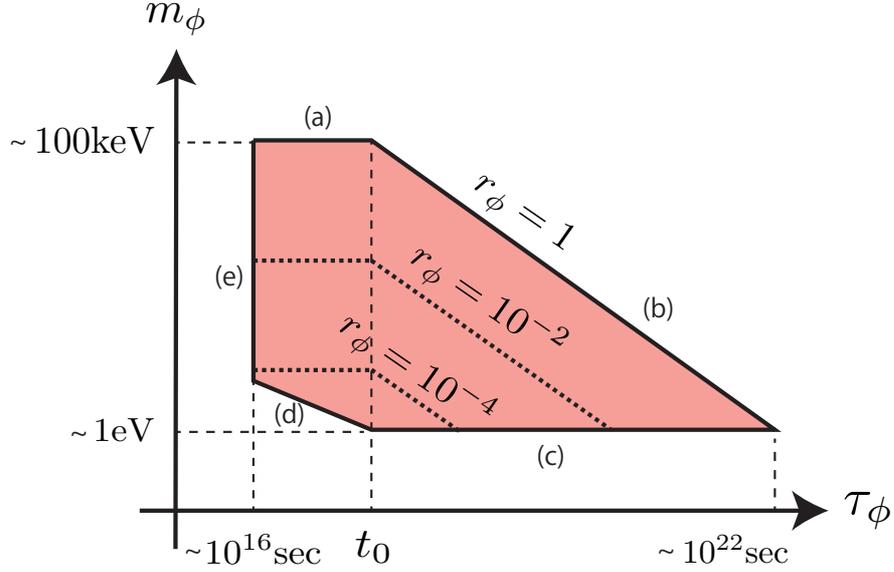}
    \caption{ The shaded parameter region on $(\tau_\phi,m_\phi)$ plane
      may explain the observed CIB excess.  The boundaries are obtained
      from the consideration about the upper bound on the dark-matter
      density (a and b), the position of the peak of the CIB spectrum
      (c and d), and the bound $E_{\rm max}<13.6\ {\rm eV}$ (e).  This
  figure should be regarded as a schematic one, and numerical values
  written in the figure are also just reference values, not exact one.
  Dotted lines are contours of $r_\phi$ to explain the CIB excess.  }
\label{fig:para}
\end{center}
\end{figure}

The resultant constraints on the parameter space is schematically
shown in Fig.~\ref{fig:para}.  Roughly, we need $10^{16}\,{\rm
  sec}\lesssim \tau_\phi\lesssim 10^{22}\,{\rm sec}$ and $2\,{\rm eV}
\lesssim m_\phi \lesssim 100\,{\rm keV}$.

\subsection{Model}

Here we briefly discuss possible models to explain the CIB excess,
although it will become clear that such models suffer from strong constraints from the CIB anisotropy measurement as shown in the next section.

One may imagine a massive pseudo-scalar particle $\phi$, which
interacts as
\begin{align}
  \mathcal L = -\frac{\phi}{M} F_{\mu\nu}\widetilde F^{\mu\nu}.
  \label{phi-gamma-gamma}
\end{align}
where $F^{\mu\nu}$ is the field strength tensor of the electromagnetic
gauge boson, and $M$ is a constant.  With such an interaction, the
lifetime of $\phi$ is given by
\begin{align}
  \tau_\phi = \left(\frac{m_\phi^3}{4\pi M^2}\right)^{-1} 
  \simeq 3\times 10^{22}\,{\rm sec}
  \times
  \left( \frac{1\,{\rm eV}}{m_\phi} \right)^3
  \left( \frac{M}{10^9\,{\rm GeV}} \right)^2.
\end{align}
Thus we need $M\lesssim 10^9$\,GeV to explain the CIB excess.\footnote
{In this case, there is an extra factor of $2$ in the photon flux
  (\ref{flux}) since two photons are produced per $\phi$ decay.  Note
  also that $m_\phi = 2E_{\rm max}$ is almost fixed by the observation
  and there is no freedom to choose $m_\phi$.  }
However, such parameter region is already excluded mainly due to too
rapid cooling of the horizontal branch stars through the Primakoff
process~(see, e.g., Ref.~\cite{Jaeckel:2010ni}).  So we do not pursue
this possibility further.

If a hidden photon as well as a new pseudo-scalar boson exist, we can
avoid the rapid cooling of the horizontal branch stars through the
Primakoff process.  Let us introduce the following interaction:
\begin{align}
  \mathcal L = -\frac{\phi}{M} F_{\mu\nu}\widetilde F_H^{\mu\nu},
  \label{L_boson}
\end{align}
where $F_H^{\mu\nu}$ is the field strength of the hidden-photon.  We
assume that there is no kinetic mixing between the photon and hidden
photon nor $\phi$-$\gamma$-$\gamma$ interaction in the form of Eq.\
\eqref{phi-gamma-gamma}.  In addition, the hidden-photon may acquire
mass through the spontaneous breaking of the hidden $U(1)$ symmetry.
If $\phi$ interacts with the standard-model particles only through the
interaction \eqref{L_boson}, Primakoff process is irrelevant.

With the interaction \eqref{L_boson}, the decay rate of $\phi$ is
given by
\begin{align}
  \Gamma(\phi\to\gamma\gamma_H)=
  \frac{m_\phi^3}{8\pi M^2}\left(1-\frac{m_\chi^2}{m_\phi^2}\right)^3 =
  \frac{E_{\rm max}^3}{\pi M^2},
\end{align}
where $\gamma_H$ denotes the hidden photon, which takes the role of $\chi$ in the previous section, and $m_\chi$ is its mass.  
The rate is independent of the mass of $\phi$ once we fix $E_{\rm max}$.  Thus the lifetime is given by
\begin{align}
  \tau_\phi \simeq 4\times10^{21}\,{\rm sec}
  \times
  \left( \frac{1\,{\rm eV}}{E_{\rm max}} \right)^3
  \left( \frac{M}{10^9\,{\rm GeV}} \right)^2,
  \label{tau_phi}
\end{align}
and it is possible to increase $m_\phi$ while fixing $E_{\rm max}$ and
$\tau_\phi$ by making the $\phi$ and $\chi$ degenerate in mass.
Looking at Fig.~\ref{fig:para}, the allowed parameter region is
$10^7\,{\rm GeV}\lesssim M \lesssim 2\times 10^9\,{\rm GeV}$ and
$2\,{\rm keV} \lesssim m_\phi \lesssim 100\,{\rm keV}$.  In this
model, there is no Primakoff production of the light boson $\phi$ and
hence the standard constraint from horizontal branch stars is not
applied to this coupling.  Still, however, there is a plasmon decay
process: $\gamma \to \gamma_H\phi$, where the photon obtains an
effective mass of the plasma frequency inside the star.  It still
significantly contributes to the energy loss of the stars as will be
shown below.

Now let us discuss the constraints on our model from stellar physics. We closely follow the discussion in Ref.~\cite{Raffelt:1996wa}.
In the electron plasma, the photon obtains an effective mass $\omega_P$, called the plasma frequency, given by\footnote{
	It applies only to the transverse photon. The dispersion relation of longitudinal photon is not a standard form
	and its behavior is more subtle. Below we only consider the transverse photon.
}
\begin{align}
	\omega_P^2 = \frac{4\alpha_e}{\pi}\int dp_e f_e\,p_e \left(v_e - \frac{v_e^3}{3}\right),
\end{align}
where $\alpha_e$ is the fine-structure constant,
$f_e$ is the electron momentum distribution function, $p_e$ and $v_e$ are the momentum and velocity of the electron.
Such a photon with a plasma mass is called plasmon.
In the non-relativistic and non-Fermi-degenerate limit, we have
\begin{align}
	\omega_P^2 \simeq \frac{4\pi \alpha_e n_e}{m_e},
\end{align}
with $n_e$ and $m_e$ being the number density and mass of the electron.
In the strongly Fermi-degenerate limit, we have
\begin{align}
	\omega_P^2 \simeq \frac{4\pi \alpha_e n_e}{E_F},
\end{align}
where $E_F$ denotes the Fermi energy.
Thus whenever there is a coupling that causes a decay $\phi\to\chi\gamma$, 
there must be a plasmon decay process $\gamma\to\phi\chi$ if $\phi$ and $\chi$ are lighter than the plasmon.
The plasmon decay rate in the model of (\ref{L_boson}) is given by
\begin{align}
	\Gamma(\gamma\to\phi\chi) \simeq \frac{\omega_P^3}{8\pi M^2} \frac{\omega_P}{E},
\end{align}
for $\omega_P \gg m_\phi, m_\chi$
where $E$ denotes the incident photon energy and the factor $\omega_P/E$ represents the time dilation effect~\cite{Raffelt:1987np}.
Note that $E > \omega_P$ is always satisfied from the dispersion relation.
Assuming that $\phi$ and $\chi$ escape from the source quickly without trapped by the plasma,
the energy loss rate per unit mass is estimated by
\begin{align}
	\epsilon = \frac{1}{\rho_s \pi^2}\int dk k^2 \frac{E}{e^{E/T}-1}\Gamma(\gamma\to\phi\chi),
\end{align}
where $\rho_s$ is the mass density of the star.
Taking the limit $\omega_p\ll T$, we obtain
\begin{align}
  \epsilon = \frac{\zeta(3)}{4\pi^3} \frac{\omega_p^4 T^3}{\rho_s M^2}\simeq
  3\times 10^{-1}\,{\rm erg/g/s}
  \times
  \left( \frac{\omega_p}{1\,{\rm keV}} \right)^4
  \left( \frac{T}{10\,{\rm keV}} \right)^3
  \left( \frac{10^4\,{\rm g/cm^3}}{\rho_s} \right)
  \left( \frac{10^9\,{\rm GeV}}{M} \right)^2.
\end{align}
Numerical values inside the parenthesis correspond to typical values for the horizontal branch  and red giant stars.
The constraint reads $\epsilon \lesssim 10\,{\rm erg/g/s}$~\cite{Raffelt:1996wa}
and hence it may be consistent with the decaying particle scenario to explain the CIB mean intensity excess for $M\sim 10^9\,$GeV.
Moreover, this constraint does not apply to relatively heavy particle: $m_\phi \sim 100$\,keV.
The observation of SN1987A also gives constraint on the extra energy loss rate as $\epsilon \lesssim 10^{19}\,{\rm erg/g/s}$.
For typical parameters of the supernova core, $\omega_p \sim 10$\,MeV~\cite{Kopf:1997mv}, 
$T\sim 30$\,MeV, $\rho_s\sim 3\times 10^{14}\,{\rm g/cm^3}$,
the energy loss rate is much smaller than the upper bound.

\section{CIB Anisotropy from Decaying Particle}
\label{sec:3}

So far we have discussed only the mean CIB intensity from decaying particles.
However, since the non-relativistic matter clumps due to the gravitational potential, photons emitted from the matter are not isotropic.
Ref.~\cite{Gong:2015hke} studied the CIB anisotropy from decaying axion-like particle
for the axion lifetime much longer than the present age of the universe.
We extend the discussion to more general case.

A useful way to discuss the anisotropy is to expand the flux with spherical harmonics $Y_{\ell m}$ as
\begin{align}
	\delta I (E,\Omega)  = I (E,\Omega) - \overline{I(E)} 
	= \sum_{\ell, m} a_{\ell m}(E)Y_{\ell m}(\Omega),
\end{align}
and define $C_\ell$ by
\begin{align}
	C_\ell (E) = \left<|a_{\ell m} (E)|^2\right> = \frac{1}{2\ell +1}\sum_{m=-\ell}^{+\ell}|a_{\ell m}(E)|^2.  \label{ClE}
\end{align}
It is calculated as~\cite{Ando:2005xg}
\begin{align}
	C_\ell(E) = E^4\,\frac{2}{\pi} \int dr W(r) \int dr' W(r') \int k^2 dk \, P_{\delta}(k; r,r') j_\ell (kr) j_\ell (kr'),  \label{Cl}
\end{align}
where $W(r)$ is defined in (\ref{Wr}), $j_\ell$ is the spherical Bessel function and the power spectrum of the matter density fluctuation $\delta(\vec x, r)$ is defined as
\begin{align}
	\left< \delta_{\vec k}(r) \delta_{\vec k'}(r') \right> = (2\pi)^3\delta^3(\vec k + \vec k') P_\delta(k; r,r').
\end{align}

As for the matter power spectrum, we should take account of the non-linear structure formation effect.
It includes the two-halo contribution at large scales and one-halo contribution at small scales.
The one-halo term and two-halo terms are written as~\cite{Cooray:2002dia}
\begin{align}
	P^{\rm 1h}_\delta (k,r=r') = \frac{1}{(\rho_m^0)^2}\int dM M^2\frac{dn(M,z)}{dM}|u_M(k)|^2,
\end{align}
and 
\begin{align}
	P^{\rm 2h}_\delta (k,r=r') =\left[\frac{1}{{\rho_m^0}} \int dM M\frac{dn(M,z)}{dM}u_M(k) b(M,z) \right]^2 P_\delta^{\rm (lin)}(k,z),
\end{align}
where $dn/dM$ denotes the comoving number density of halo with mass of $M$, $\rho_m^0$ is the present matter energy density,
$u_M(k)$ denotes the Fourier transform of the density profile of each halo~\cite{Cooray:2002dia},
and $P_\delta^{\rm (lin)}(k)$ is the linear matter density perturbation with $b(M,z)$ being the linear halo bias~\cite{Eisenstein:1997jh}.  
In the numerical calculation, we use the Sheth-Tormen functional form for $dn/dM$~\cite{Sheth:1999mn}
with the Planck best-fit cosmological parameters~\cite{Ade:2015xua}.
We take the lower cutoff of the halo mass $M$ to be one solar mass in the numerical calculation,
but the result is not sensitive to this choice unless the cutoff mass is extraordinarily large.
Also we should take the effect of Doppler broadening of the photon spectrum into account.
We simply take the injection photon energy spectrum to be the box shape:
\begin{align}
	\frac{dN_\gamma}{dE'} = \begin{cases}
		(v E_{\rm max})^{-1} & {\rm for}~|E' - E_{\rm max}| < v E'/2 \\
		0 & {\rm otherwise}
	\end{cases},
	\label{box}
\end{align}
where $v$ denotes the typical velocity dispersion of $\phi$ in the halo.\footnote{
	Actually $v$ depends on the halo mass and hence it should be inside the $M$ integral in $P_\delta$.
	However, for our purpose to show that the anisotropy spectrum significantly exceeds the observed spectrum,
	it is enough to fix $v$ $(\sim 10^{-3})$ as a representative value in the large halo.
	Since $v$ is expected to be smaller in small halos and it enhances the anisotropy spectrum,
	taking account of precise halo dependence of $v$ would make the discrepancy even larger.
}

We evaluate (\ref{Cl}) by the following approximation.
First note that the $k$ integration in (\ref{Cl}) depends on the shape of $P_\delta(k)$.
We are interested in the wave number of $k \gtrsim 0.1\,{\rm Mpc}^{-1}$. 
In this range, it has nearly flat spectrum below a critical wavenumber $k < k_c$,
and it is proportional to $\sim k^{-2}$ for $k > k_c$ where $k_c$ depends on the redshift $z$.
In such a case, we can approximate the $k$ integration in (\ref{Cl}) as
\begin{align}
	\frac{2}{\pi}\int k^2 dk \, P_{\delta}(k; r,r') j_\ell (kr) j_\ell (kr')\sim
	\begin{cases}
	 \displaystyle \frac{1}{r^2 \Delta r_\ell} P(k=k_*; r) & {\rm for}~~|r-r'| <\Delta r_\ell,\\
	 0 & {\rm otherwise}
	 \end{cases},
\end{align}
where we have defined $\Delta r_\ell = 1/k_*$ and
\begin{align}
	k_* = {\rm max} \left[ \frac{\ell}{r},~k_c \right].
\end{align}
It is consistent with the approximation done in Ref.~\cite{Ando:2005xg} for $\ell/r > k_c$.
Second, recall that the function $W(r)$ contains a box-shape function (\ref{box}) whose width is given by $\Delta r_v \equiv v(1+z)/H(z)$
in terms of the width of the comoving distance $r$. 
Hence the effective integration range depends on whether $\Delta r_v > \Delta r_\ell$ or $\Delta r_v < \Delta r_\ell$.
If $\Delta r_v > \Delta r_\ell$, we find
\begin{align}
	C_\ell(E) \sim \left(\frac{c}{4\pi}\frac{n_\phi(z)a^3(z)}{\tau_\phi} \frac{E}{H(z)} \right)^2 \frac{\Delta r_\ell}{\Delta r_v}
	\frac{2\pi^2}{r^2(z) k_*^2} \mathcal P_{\delta}(k=k_*; r(z)),
	\label{Cl1}
\end{align}
where $z$ is fixed from $E(1+z)=E_{\rm max}$.
If, on the other hand, $\Delta r_v < \Delta r_\ell$, we obtain
\begin{align}
	C_\ell(E) \sim \left(\frac{c}{4\pi}\frac{n_\phi(z)a^3(z)}{\tau_\phi}\frac{E}{H(z)} \right)^2
	\frac{2\pi^2}{r^2(z) k_*^2} \mathcal P_{\delta}(k=k_*; r(z)).
	\label{Cl2}
\end{align}
Here $\mathcal P_\delta(k)\equiv (k^3/2\pi^2) P_\delta(k)$ denotes the dimensionless power spectrum.
The quantity in the first parenthesis represents the mean intensity (\ref{flux}).

\begin{figure}[t]
\begin{center}
\includegraphics[scale=1.2]{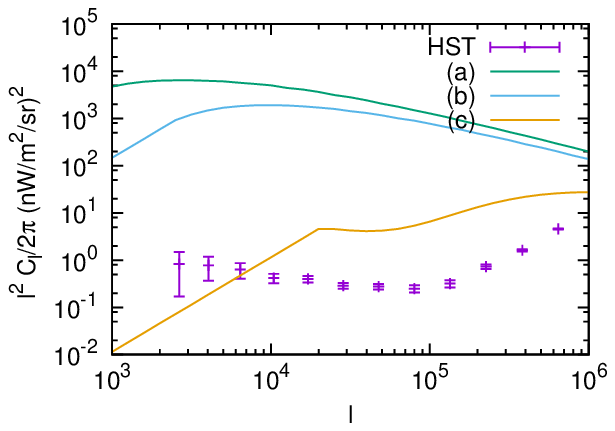}
\includegraphics[scale=1.2]{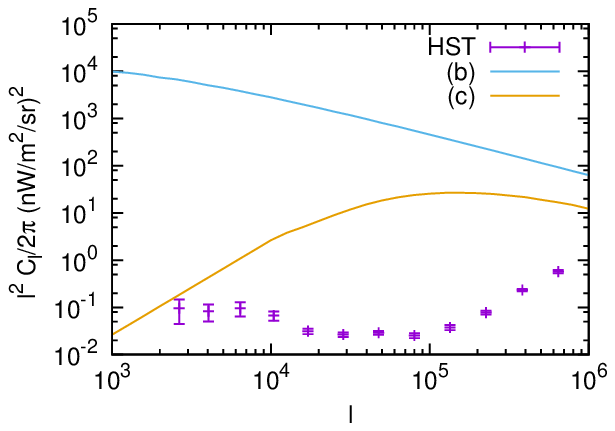}
\caption{
The CIB anisotropy spectrum from decaying particle at the observation wavelength $\lambda= 1.6\,\mu$m (left)
and  $\lambda= 0.85\,\mu$m (right).
Data points correspond to the HST measurement at the same wavelength.
The solid lines show the prediction from decaying particle for the parameter set same as those of Fig.~\ref{fig:spec}.
(a) $E_{\rm max}=1$\,eV, $m_\phi/r_\phi=3$\,eV, $\tau_\phi=2\times10^{22}$\,sec,
(b) $E_{\rm max}=1.5$\,eV, $m_\phi/r_\phi=90$\,keV, $\tau_\phi=2.5\times10^{17}$\,sec,
(c) $E_{\rm max}=8$\,eV, $m_\phi/r_\phi=80$\,keV, $\tau_\phi=1\times10^{16}$\,sec. 
}
\label{fig:aniso}
\end{center}
\end{figure}

Fig.~\ref{fig:aniso} shows the CIB anisotropy spectrum from decaying particle calculated by (\ref{Cl1}) and (\ref{Cl2})
at the observation wavelength $\lambda= 1.6\,\mu$m (left) and  $\lambda= 0.85\,\mu$m (right).
Data points correspond to the HST measurement at the same wavelength~\cite{Mitchell-Wynne:2015rha}.
The solid lines show the prediction from decaying particle for the parameter set same as those of Fig.~\ref{fig:spec}, i.e.,
(a) $E_{\rm max}=1$\,eV, $m_\phi/r_\phi=3$\,eV, $\tau_\phi=2\times10^{22}$\,sec,
(b) $E_{\rm max}=1.5$\,eV, $m_\phi/r_\phi=90$\,keV, $\tau_\phi=2.5\times10^{17}$\,sec,
(c) $E_{\rm max}=8$\,eV, $m_\phi/r_\phi=80$\,keV, $\tau_\phi=1\times10^{16}$\,sec. 
For the parameter set (a), there is no contribution to the 0.85\,$\mu$m background light at present. 
We have also taken $v=4\times 10^{-3}$ as a representative value in the largest halo ($M \sim 10^{14}$ times solar mass).
It is clearly seen that the predictions exceed the results of HST measurement by orders of magnitude.
Although not shown in the figure, it also exceeds the data points from CIBER anisotropy measurement~\cite{Zemcov:2014eca}.
In the case of (c) the discrepancy is rather mild because the observed photons at the wavelength $\lambda= 1.6\,\mu$m
comes from the high redshift due to large $E_{\rm max}$ and the structure formation at the high redshift is relatively suppressed.
Even in such a case the predicted anisotropy is far above the observational bound for $\lambda= 0.85\,\mu$m.
For larger $E_{\rm max}$ the redshift of the emission epoch becomes larger and the anisotropy is suppressed,
but $E_{\rm max}$ is bounded as $E_{\rm max} < 13.6$\,eV so that the $\phi$ decay does not affect reionization history too much
as mentioned in the previous section.\footnote{
	For $10.2$\,eV $< E_{\rm max} < 13.6$\,eV, there is another complexity due to the Lyman $\alpha$ absorption,
	hence we do not consider this energy range.
}
Therefore we conclude that it is difficult for a decaying particle to explain the CIB spectrum excess in the mean intensity
without producing too much anisotropy.

\section{Conclusions and Discussion}
\label{sec:4}

In this paper we considered a possibility that the recent observation of the CIB mean intensity
excess by the CIBER experiment as well as the AKARI and IRTS may be explained by decaying light particles.
Indeed it is possible to fit the observed CIB spectrum of the mean intensity,
but it necessarily predicts too much CIB anisotropy, which contradicts with the anisotropy measurements by the HST and CIBER.

This indicates that the excess CIB radiation may have some astrophysical origin. 
Here we briefly comment on severe problems of the excess of the
diffuse CIB radiation with observed TeV gamma-rays. A TeV gamma-ray
scatters off the CIB photon and produces electron and positron through
$\gamma + \gamma \to e^- + e^+$. Then the TeV gamma-ray cannot travel
from a high-redshifted object such as Active Galactic Nuclei or
blazars (with their redshift $z \gg 0.1$).
The HESS collaboration reported the upper bound on the flux of the CIB
radiation~\cite{Aharonian:2005gh} to be consistent with the TeV
gamma-ray observations. The value of the CIB radiation flux reported
by the CIBER collaboration exceeds this upper bound.  In this
circumstance, there may exist a non-standard mechanism to solve this
problem.  For example, the TeV gamma-rays could have origins in
line-of-sight cosmic-ray proton interactions with background photons;
their continuous gamma-ray productions during their travel may be the
source of TeV gamma-rays~\cite{Essey:2010er}. This is different from a
normal scenario in which the TeV gamma-rays are produced at the
sources of the proton acceleration.  In such a scenario, the TeV
gamma-rays produced at low redshift $z \ll 0.1$ are not absorbed by
the CIB radiation and can be observed in the Milky-Way (MW) Galaxy.
As another possibility, we may consider models with axion or
axion-like particles (ALPs) which mix with the photon when magnetic
field exists~\cite{Raffelt:1987im}.  We can expect possible
oscillations of the TeV gamma-rays into ALPs at around sources
utilizing magnetic field.  ALPs are not absorbed during their travel
from the distant source to the MW Galaxy. We can observe the TeV
gamma-rays oscillated-back from ALPs within the MW
Galaxy~\cite{Raffelt:1996wa,Hooper:2007bq,SanchezConde:2009wu,Kohri:2017ljt}.

\section*{Acknowledgments}

This work is supported by JSPS KAKENHI Grant Numbers
26247042 (K.N. and K.K.), JP15H05889 (K.K.), JP16H0877 (K.K.), 
JP17H01131 (K.K.), 
26400239 (T.M.), 16H06490 (T.M.), 
26800121 (K.N.), 26104009 (K.N.), and JP15H05888 (K.N.).



\end{document}